\documentclass{article}
\usepackage{spconf,amsmath,epsfig}

\title{Hybrid-ARQ in Multihop Networks with Opportunistic Relay Selection}
\name{Caleb K. Lo, Robert W. Heath, Jr. and Sriram Vishwanath \thanks{Caleb K. Lo was supported by a Microelectronics and Computer Development (MCD) Fellowship and a Thrust 2000 Endowed Graduate Fellowship through The University of Texas at Austin.  Robert Heath was supported in part by the National Science Foundation under grant CNS-626797 and the DARPA IT-MANET program, Grant W911NF-07-1-0028.  Sriram Vishwanath was supported by the National Science Foundation under grants CCF-055274, CCF-0448181, CNS-0615061 and CNS-0626903.}}
\address{The University of Texas at Austin, Austin, Texas 78712 \\ Email: \{clo, rheath, sriram\}@ece.utexas.edu}
\begin{document}
\topmargin=0mm
% \setlength{\evensidemargin}{-6.2mm}
% \setlength{\oddsidemargin}{-6.2mm}\setlength{\textwidth}{178mm}
% \setlength{\topmargin}{0in} \setlength{\columnsep}{6mm}
% \setlength{\textheight}{229mm } \pagestyle{empty}
% \marginsize{0.75in}{0.75in}{1in}{1in}
% \ninept
%
\maketitle
\begin{abstract}
This paper develops a contention-based opportunistic feedback technique towards relay selection in a dense wireless network.  This technique enables the forwarding of additional parity information from the selected relay to the destination.  For a given network, the effects of varying key parameters such as the feedback probability are presented and discussed.  A primary advantage of the proposed technique is that relay selection can be performed in a distributed way.  Simulation results find its performance to closely match that of centralized schemes that utilize GPS information, unlike the proposed method.  The proposed relay selection method is also found to achieve throughput gains over a point-to-point transmission strategy.
\end{abstract}
\section{Introduction}
Mesh networks are integral to the operation of next-generation wireless systems.  One of the key aspects of mesh networks is their ability to support multihop signaling, where intermediate nodes can act as relays by forwarding a message from a source to a distant destination.  Message forwarding occurs over inherently unstable wireless links; thus, throughput and reliability are decreased by packet outages.  Hybrid automatic-repeat-request (ARQ) has been proposed as an enabling method for satisfying quality-of-service (QoS) constraints in wireless systems.  Hybrid-ARQ methods are particularly useful in mesh networks, where relays that are closer to the destination than the source can forward additional parity information to the destination if the destination detects uncorrectable packet errors \cite{ZhaVal:PracRelaNetwGene:Jan:05}.  This decreases the number of retransmission cycles that are needed for decoding the source message.

Two-hop networks are especially useful for improving coverage and throughput in cellular systems.  In a two-hop network, the source can select either a single relay or multiple relays to forward its message to the destination.  There has been significant prior work on multiple relay selection \cite{LanWor:DistSpacTimeCode:Oct:03} and single relay selection \cite{CheSerETAL:DistPoweAlloPara:Nov:05,LinErk:RelaSearAlgoCode:Nov:05,SadHanETAL:DistRelaAssiAlgo:Jun:06,ZhaAdvETAL:ImprAmplForwRela:Jul:06,SreYanETAL:RelaSeleStraCell:Oct:03,BleKhiETAL:SimpCoopDiveMeth:Mar:06,LuoBluETAL:ApprCoopMultAnte:Sep:04,ZhaVal:PracRelaNetwGene:Jan:05}.  In the work most closely related to this paper, \cite{ZhaVal:PracRelaNetwGene:Jan:05}, GPS information is used to select the closest decoding relay to the destination to forward parity information.  This selection method optimizes the average SNR at the destination, but the necessity of using GPS information in the selection process makes practical implementation difficult.  Further, global network information is required at all nodes which becomes more difficult to obtain and store as the number of nodes increases; a more decentralized method for relay selection would be preferable.

In this paper, we propose a decentralized relay selection approach that relies on random access-based feedback to the source.  Relay selection is based on $\textit{opportunistic}$ $\textit{feedback}$ \cite{TanHea:OppoFeedDownMult:Oct:05}, which is designed for user selection in a downlink wireless system.  In our approach, the source uses ``Hello'' messages from the relays to select a relay to forward parity information to the destination if it detects an uncorrectable packet error.  The ``Hello'' message feedback is controlled by factors including the relay channel gain to the destination.  We present and discuss the effects on system performance of varying key parameters such as the feedback probability and channel threshold.  Our approach significantly outperforms a point-to-point hybrid-ARQ strategy where the source forwards the parity information to the destination.  Also, our strategy yields throughputs that are very close to those yielded by the GPS-based strategy in \cite{ZhaVal:PracRelaNetwGene:Jan:05}.  This further demonstrates the utility of decentralized relay selection algorithms in dense networks.

We use boldface notation for vectors.  SNR represents the signal-to-noise ratio.  $\|\mathcal{A}\|$ denotes the cardinality of a set $\mathcal{A}$.

\section{System Model}
\begin{figure}[tb]
\begin{center}
\includegraphics[width=3.0in]{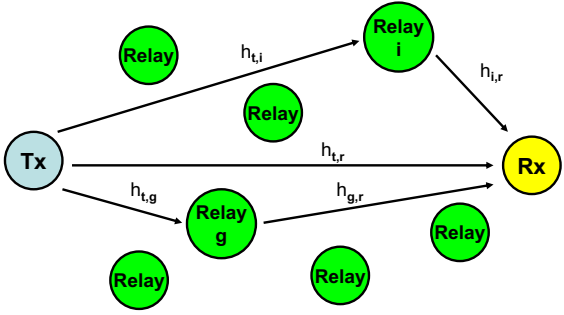}
\end{center}
\caption{Relay network.}
\label{system-model}
\end{figure}

Consider the setup in Fig. \ref{system-model}.  There are $K_r$ relays that are interspersed in the region between the source and the destination.  We adopt the system constraints in \cite{ZhaVal:PracRelaNetwGene:Jan:05}, so each relay operates in a half-duplex mode and is equipped with a single antenna.  In particular, when either the source, one of the relays, or the destination sends a message, all of the other nodes are in a receiving mode.

% Below is an example of how to insert images. Delete the ``\vspace'' line,
% uncomment the preceding line ``\centerline...'' and replace ``imageX.ps''
% with a suitable PostScript file name.
% -------------------------------------------------------------------------

Transmission occurs over a set of time slots $\{t_1,...,t_m\}$ which are of equal duration.  Initially, the source has a k-bit message $\textbf{w}$ that is encoded as an n-bit codeword $\textbf{x}$.  We adopt the ARQ/FEC protocol in \cite{Hag:RateCompPuncConv:Apr:88}, so the source chooses code rates $\{R_1,R_2,...,R_m\}$ from a rate-compatible punctured convolutional code (RCPC) family, and $R_1 > R_2 > \cdots > R_m$.  The rate-$R_m$ code is the mother code of the RCPC family.

Before $t_1$, the source and destination perform RTS/CTS-based handshaking to achieve synchronization.  During $t_1$, the source transmits a subset $\textbf{x}_1$ of the bits in $\textbf{x}$ such that $\textbf{x}_1$ forms a codeword from the rate-$R_1$ code.  The destination observes
\begin{equation}
\textbf{y}_{r,1} = h_{t,r}\textbf{x}_1 + \textbf{n}_r
\end{equation}
while relay $i \in \{1,2,...,K_r\}$ observes
\begin{equation}
\textbf{y}_{i,1} = h_{t,i}\textbf{x}_1 + \textbf{n}_i.
\end{equation}
Here, $h_{t,i}$ represents a Rayleigh fading coefficient for the channel between the source and node $i$, while $\textbf{n}_i$ represents additive white Gaussian noise with variance $N_0$ at node $i$.

We assume that all fading coefficients are constant over a time slot and vary from slot to slot, which is a valid assumption given that each time slot is less than the channel coherence time.  It is also assumed that fading and additive noise are independent across the nodes, which are valid assumptions given that node separation is greater than the channel coherence distance.  It is also assumed that all nodes have no knowledge of fading coefficients and must learn them via training data at the beginning of each packet transmission.

The destination attempts to decode $\textbf{y}_{r,1}$.  If decoding is successful, the destination broadcasts an ACK message to all of the relays and the source.  If decoding is unsuccessful, the destination broadcasts a NACK message to all of the relays and the source.  The challenge for the source is to select one of the relays to forward additional parity information that will assist the destination in recovering $\textbf{w}$.  We now describe our method for relay selection.

\section{Opportunistic Relay Selection}
\begin{figure}[tb]
\begin{center}
\includegraphics[width=3.0in]{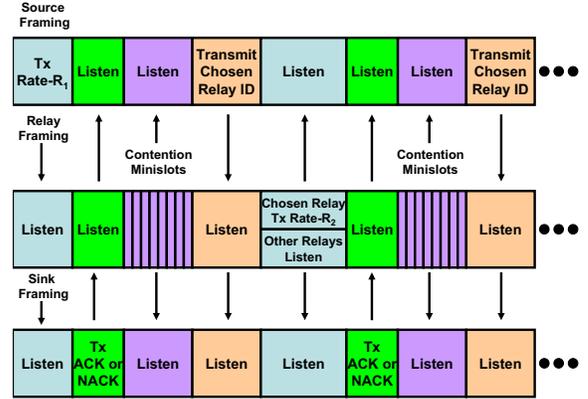}
\end{center}
\caption{Framing structure for proposed selection strategy.}
\label{framing-structure}
\end{figure}

We modify the opportunistic feedback approach in \cite{TanHea:OppoFeedDownMult:Oct:05} to select one of the relays.  The framing structure for our algorithm is shown in Fig. \ref{framing-structure}.  In Fig. \ref{framing-structure} it is assumed that a NACK is sent after each packet transmission, which automatically starts the relay contention process.  Let $\mathcal{R}_{sel}$ denote the set of relays that can participate in the relay selection process, where relay $i \in \mathcal{R}_{sel}$ has both recovered $\textbf{w}$ and has a channel gain to the destination $|h_{i,r}|^2$ that is above a threshold $\eta_{opp}$.  Each relay $i$ can determine $|h_{i,r}|^2$ by listening to the destination's ACK or NACK message after a packet transmission; the ACK or NACK message is embedded in a packet that contains training data.  All relays in $\mathcal{R}_{sel}$ are allocated the same $K$ minislots for feedback to the source.

During minislot $b$, each relay $i \in \mathcal{R}_{sel}$ will send a ``Hello'' message to the source with probability $p_i$.  Successful contention occurs during minislot $b$ if exactly one relay $i \in \mathcal{R}_{sel}$ sends a ``Hello'' message to the source.  If relays $s, t \in \mathcal{R}_{sel}$ send ``Hello'' messages during minislot $b$ and $s \neq t$, a collision occurs and the source discards all received ``Hello'' messages.  After minislot $K$ has been completed, the source determines if successful contention has occurred for at least one minislot $b$.  If so, the source randomly selects a relay $i_{t}$ that has successfully sent a ``Hello'' message to it; otherwise the source will transmit during $t_2$.

During $t_2$, relay $i_{t}$ (or the source) transmits a subset $\textbf{x}_2$ of the bits in $\textbf{x}$ such that $\textbf{x}_1 \cup \textbf{x}_2$ forms a codeword from the rate-$R_2$ code.  This means that the destination should not discard $\textbf{y}_{r,1}$ after $t_1$; instead, it should combine $\textbf{y}_{r,1}$ with
\begin{equation}
\textbf{y}_{r,2} = h_{i_{t},r}\textbf{x}_2 + \textbf{n}_r
\end{equation}
and attempt to decode $\textbf{y}_{r,1} \cup \textbf{y}_{r,2}$ based on the rate-$R_2$ code.  If decoding at the destination is unsuccessful, the destination broadcasts another NACK message to all of the relays and the source, and then we repeat the relay contention process to select another relay to transmit during $t_3$.  This process repeats until the destination successfully recovers $\textbf{w}$ or the rate-$R_m$ code has been used without successful decoding.

To compute the throughput of this scheme, we use (16) from \cite{Hag:RateCompPuncConv:Apr:88}
\begin{equation}\label{hagenauer-throughput}
R_{avg} = \frac{k}{n+M}\cdot \frac{P}{P+l_{AV}}
\end{equation}
where $l_{AV}$ is the average number of additionally transmitted bits per $P$ information bits, $P$ is the puncturing period of the RCPC family and $M$ is the memory of the mother code.

Since relay $i$ can determine $|h_{i,r}|^2$, we could have modified our protocol to have all of the decoding relays perform distributed beamforming.  Distributed beamforming is difficult to implement in practice, though, since the oscillators in distinct nodes are not necessarily synchronized and are subject to phase noise.  We could have also modified our protocol to have all of the decoding relays forward their parity information using orthogonal time slots, but this would tax system resources as $\|\mathcal{R}_{sel}\|$ grows large.

\section{Performance Impact of Varying System Parameters}\label{perf-impact}
By optimizing parameters such as the relay-to-source feedback probability $p_i$ and the relay-to-destination channel threshold $\eta_{opp}$, we can maximize the throughput for our approach.  A joint optimization of these parameters is fairly difficult, though, so in this section we provide some insight as to how each parameter individually affects the throughput.

For simulation purposes, we employ the path loss model described in \cite{ZhaVal:PracRelaNetwGene:Jan:05}; thus, the received energy at node $i$ is
\begin{eqnarray}
\mathcal{E}_i & = & |h_{b,i}|^2\mathcal{E}_{x_1} \\
& = & (\lambda_c/4\pi d_0)^2(d_{b,i}/d_0)^{-\mu}\mathcal{E}_{x_1}
\end{eqnarray}
where $\mathcal{E}_{x_1}$ is the transmitted energy in $x_1$.  Here, $\lambda_c$ is the carrier wavelength, $d_0$ is a reference distance, $d_{b,i}$ is the distance between transmitting node $b$ and receiving node $i$, and $\mu$ is a path loss exponent.

We adopt similar simulation parameters as those in \cite{ZhaVal:PracRelaNetwGene:Jan:05}.  Here, we employ a carrier frequency $f_c$ = 2.4GHz, $d_0$ = 1m, $d_{t,r}$ = 100m and $\mu$ = 3, where $d_{t,r}$ is the distance between the source and the destination.  We then uniformly distribute $K_r = 20$ relays in the region between the source and the destination such that each relay $i$ is $d_{i,r} < d_{t,r}$ units from the destination.  We also use the WiMAX signaling bandwidth, which is roughly 9 MHz \cite{WireMANWorkGrp}; given a noise floor of -204 dB/Hz this yields a noise value $N_0 = -134$ dB.  BPSK modulation is used for all packet transmissions, and all of the relays and the destination use ML decoding.

We employ the codes of rates $\{4/5, 2/3, 4/7, 1/2, 1/3\}$ from the $M = 6$ RCPC family in \cite{Hag:RateCompPuncConv:Apr:88}.  We perform concatenated coding, where the outer code is a (255, 239) Reed-Solomon code with symbols from $GF(2^8)$; this code can correct at most 8 errors.  The mother code for the RCPC family is a rate-1/3 convolutional code with constraint length 7 and generator polynomial (145 171 133) in octal notation.

\begin{figure}[tb]
\begin{center}
\includegraphics[width=3.0in]{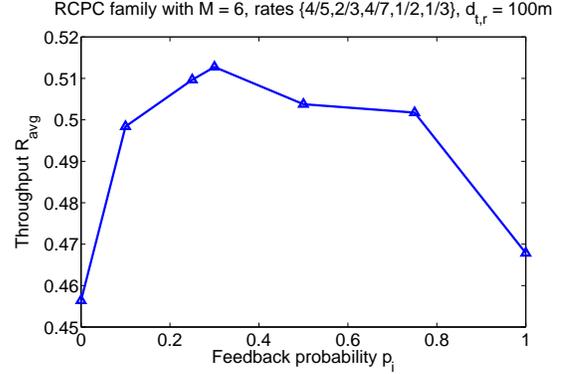}
\end{center}
\caption{Throughput as a function of feedback probability.}
\label{fbk-prob}
\end{figure}

Fig. \ref{fbk-prob} shows how the throughput $R_{avg}$ yielded by our selection approach varies with the feedback probability $p_i$.  Here we fix $K = 10$ minislots and set the channel feedback threshold $\eta_{opp} = -91 dB$.  The average received SNR at the destination is 2 dB.  We see that the throughput is maximized around $p_i = 0.3$.

The observed throughput performance has a nice intuitive explanation.  For large values of the feedback probability $p_i$, each relay node $i \in \mathcal{R}_{sel}$ is more likely to send a ``Hello'' message to the source during each minislot $b$, which increases the likelihood of a collision during minislot $b$; this also increases the likelihood that no relays will be selected during the entire contention period and that the source will end up forwarding the next set of parity bits to the destination.  For small values of the feedback probability $p_i$, each relay node $i \in \mathcal{R}_{sel}$ is less likely to send a ``Hello'' message to the source during each minislot $b$, which decreases the likelihood of successful contention in minislot $b$; again, this increases the likelihood that the source will end up forwarding the next set of parity bits to the destination.

\begin{figure}[tb]
\begin{center}
\includegraphics[width=3.0in]{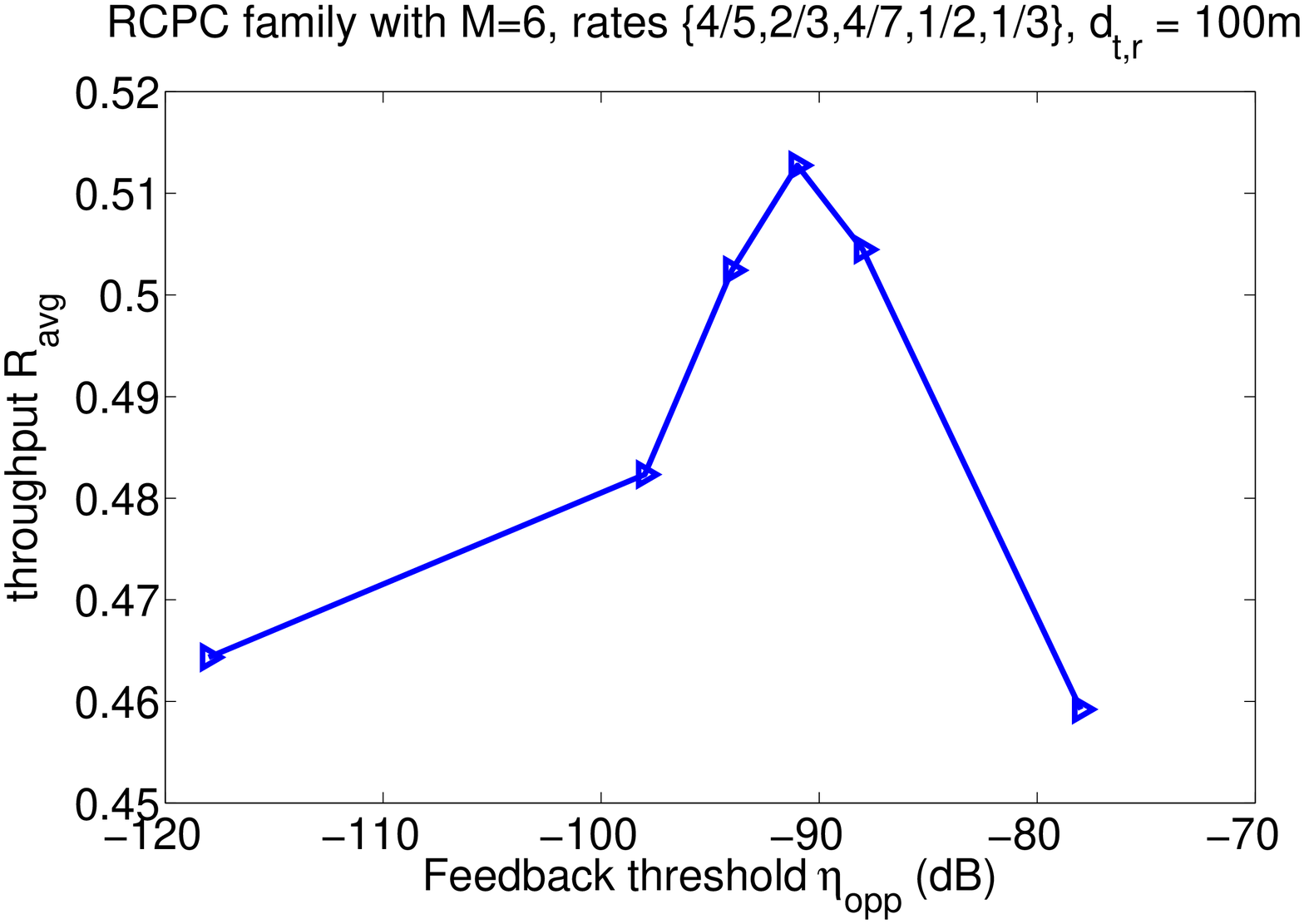}
\end{center}
\caption{Throughput as a function of feedback threshold.}
\label{eta-thresh}
\end{figure}

Fig. \ref{eta-thresh} shows how the throughput $R_{avg}$ yielded by our selection approach varies with the channel feedback threshold $\eta_{opp}$.  Here we fix $K = 10$ minislots and set the feedback probability $p_i = 0.1$.  The average received SNR at the destination is 2 dB.  We see that the throughput is maximized around $\eta_{opp} = -91 dB$.  The observed performance can be intuitively explained as follows.  For large values of the feedback threshold $\eta_{opp}$, $\|\mathcal{R}_{sel}\|$ is small, which decreases the likelihood of successful contention in minislot $b$.  For small values of the feedback threshold $\eta_{opp}$, $\|\mathcal{R}_{sel}\|$ is large, which increases the likelihood of a collision in minislot $b$.

\section{Throughput Comparison with GPS-based Strategy}
In this section we compare the throughput of our proposed strategy with the throughput of the GPS-based HARBINGER approach in \cite{ZhaVal:PracRelaNetwGene:Jan:05}.  We also consider the throughput of a point-to-point transmission strategy where the source always forwards additional parity bits to the destination.  We set $\eta_{opp} = -91 dB$, $p_i$ = 0.3, and $K = 10$ minislots; the other simulation parameters are the same as in Section \ref{perf-impact}.

\begin{figure}[tb]
\begin{center}
\includegraphics[width=3.0in]{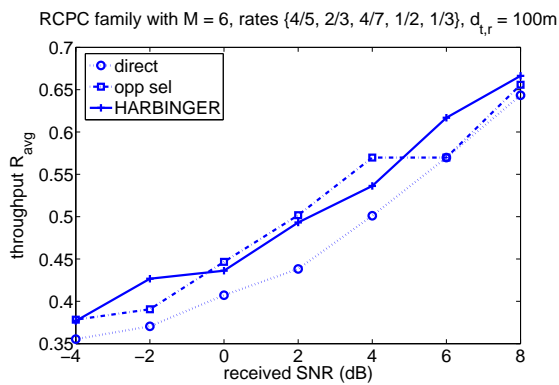}
\end{center}
\caption{Comparison with GPS-based strategy in \cite{ZhaVal:PracRelaNetwGene:Jan:05}.}
\label{throughput}
\end{figure}

We see in Fig. \ref{throughput} that our proposed approach yields results that are comparable to those yielded by the HARBINGER approach; in some cases, the decentralized strategy outperforms the HARBINGER approach.  This demonstrates that random access-based schemes can yield good performance.  Recall that the HARBINGER method optimizes the average received SNR at the destination by selecting the closest decoding relay to the destination to forward parity information.  This method, though, does not necessarily select the decoding relay that would yield the highest instantaneous received SNR at the destination.  Thus, the proposed approach can outperform the HARBINGER method in some cases.

\end{document}